\documentclass[a4paper,12pt]{amsart}

\usepackage{kantlipsum} 
\calclayout
\usepackage{setspace}

\usepackage{tikz}
\usetikzlibrary{positioning,shadows,arrows}

\usepackage{enumerate}
\usepackage{enumitem}
\usepackage{amsmath}    
\usepackage{amssymb}    
\usepackage{bm}         
\usepackage{graphicx}   
\usepackage{color}      
\usepackage{hyperref}   
\usepackage{natbib}    
\usepackage{epsf}
\usepackage{lscape}
\bibpunct{(}{)}{;}{a}{,}{,}
\usepackage{caption}

\theoremstyle{definition}

 \theoremstyle{remark}

\numberwithin{equation}{section}
\usepackage{array}
\usepackage{layout}

\usepackage[norelsize]{algorithm2e}

\begin{document}



\title[]{\textbf{A Bayesian approach for small area population estimates using multiple administrative records}}


\author[]{Jairo F\'uquene$^{\dagger}$, Andryu Mendoza$^{\ddagger}$, Cesar Cristancho$^{\ddagger}$, Mariana Ospina$^{\ddagger}$. \\
\small Consultant for National Administrative Department of Statistics on the data for health initiative, Colombia$^{\dagger}$.\\
\small Population Projections Division. National Administrative Department of Statistics, Colombia$^{\ddagger}.$}

\date{}

\maketitle

\begin{center}
\textbf{Abstract}

\end{center}
Small area population estimates are useful for decision making in the private and public sectors. However, in small areas (i.e., those that are difficult to reach and with small population sizes) computing demographic quantities is complicated. Bayesian methods are an alternative for demographic population estimates which uses data from multiple administrative records. In this paper we explore a Bayesian approach which is simple and flexible and represents
an alternative procedure for base population estimates particularly powerful for intercensal periods. The applicability of the methodological procedure is illustrated using population pyramids in the municipality of Jamund\'i in Colombia.
\vspace*{.3in}



\section{Introduction}

Population  estimates in the small areas of Colombia are useful for economic, political and social decision making. Those small areas include areas of difficult to access  (e.g., indigenous populations), areas where there is an armed conflict or municipalities with population sizes of less than 100,000 habitants. For large areas, which include departments and municipalities with high density population (typically more than 100,000 habitants) and some special areas such as metropolitan areas, the National Administrative Department of Statistics (DANE, by its Spanish acronym) uses the cohort-component method to compute population projections given by

\begin{align}
q^{c}_{l,t}=q^{c}_{l,t-1}+N_{l,t}-D_{l,t}+M_{l,t}-E_{l,t},
\end{align}

where $q^{c}_{l,t}$ is the population over time $t$, $q^{c}_{l,t-1}$ is the population over time $t-1$ and where $N_{l,t}$, $D_{l,t}$, $M_{l,t}$ and $E_{l,t}$ denote births, deaths, immigration and emigration over time $t$ and the classification age group $l$ (see for instance, \cite{preston2000demography} and \cite{wachter2014essential}). For small area population projections DANE uses the cohort relation method which implements demographic variables such as total births, survival rates, municipal  differential growth rates and five year age groups \citep{colombiaPROY}. This method is a straightforward procedure for population projection estimates however for long intercensal periods this alternative may produce biased population indicators. \\

As an alternative procedure to direct methods in demographics (e.g., the cohort-component method) probabilistic methods have been proposed with the advantage of producing
estimates and projections with their respective reliability. Demographic literature where the focus is estimation
include \cite{smith2013practitioner} and \cite{swanson2012subnational}. Bayesian methods represent an alternative probabilistic procedure for estimates and projections of population indicators.  They can be applied to either complex or simpler demographic models with the advantage of using the auxiliary information available regardless of its heterogeneity. Bayesian models in demographic can be displayed in two scenarios: 1) Bayesian methods
that use demographic variables based in the cohort-component method and incorporate prior information on traditional
demographic models (e.g., double-logistic
functions implemented in United \cite{nations2015world} or models for mortality rates proposed by \cite{lee1992modeling}) to produce indicators such as life expectancy,
mortality rates, fertility rates and size populations, and 2) Bayesian methods that use heterogeneous auxiliary information obtained through administrative records in a hierarchical (i.e., multi-level) way for producing estimates of population projections. In the first scenario, \cite{ alkema2011probabilistic} and \cite{ raftery2013bayesian} suggest estimating the fertility rate and calculating life expectancy through Bayesian models that use a double logistic function. Through the transformation of Bayesian estimates on life expectancy and fertility rates and with the help of previously defined demographic indicators, \cite{ raftery2014bayesian} estimate population projections using the cohort-component method. In the second scenario, \cite{alkema2012estimating} and \cite{alexander2018global} incorporate multiple data sources in Bayesian models to estimate fertility rates and mortality rates.
\cite{bryant2013bayesian} and \cite{bryant2015bayesian} proposed hierarchical Bayesian models for inferring demographic quantities
using information from unreliable data of administrative records. \cite{daponte1997bayesian} proposed a Bayesian model for projecting
the population size of Iraqi Kurdish population. \cite{wheldon2015bayesian} and \cite{wheldon2016bayesian} estimate simultaneously fertility, mortality, migration and size population using Bayesian models with multiple data sources.\\

This paper introduces an alternative procedure to estimate demographic base populations using information from an integrated administrative record comprised of different heterogeneous sources and prior information based on population projection estimates.
 The proposed procedure present a remarkable practical appealing in the estimation of population projections and demographic indicators. Firstly, the compensation equation of the components method is maintained and therefore the usual demographic variables are part of the analysis. Secondly, estimations of population projections can be obtained with their respective credibility interval and the uncertainty of these estimates can be established using the Baye's rule. Additionally, the use of administrative records in the Bayesian model is the greatest advantage and it opens the door to the possibility of estimating base population for long intercensal periods. One potential criticism of
 Bayesian methods is the incorporation of prior information in the Bayesian model which may be empirical or objective (based on data or robust models) and subjective (based on the expert opinion). However, for the proposed model it is possible to incorporate the prior information in the model using the demographer experience and the population projection estimates. Bayesian models may also be computationally expensive due to the use of Markov Chain Monte Carlo methods (\cite{brooks2011handbook}). However, in this proposal this is not a practical inconvenience since all calculations can be made in a closed manner without the need of computational approaches. The rest of the paper is organized as follow. In Section \ref{ssec:dos}, the methodological framework of this research is presented. In Section \ref{ssec:tres}, the proposed model for computing the base population estimates in a municipality of Colombia is used. Finally, in Section \ref{ssec:cuatro} the conclusions and remarks of this research are presented.

\section{Methodology}
\label{ssec:dos}

The main goal
of this paper is estimating the base population in small areas for long intercensal periods.
As an innovation in relation to proposed methodologies given in \cite{alkema2012estimating, alexander2018global, bryant2013bayesian, bryant2015bayesian, daponte1997bayesian,
wheldon2015bayesian, wheldon2016bayesian},
we use an integrated administrative record incorporating multiple heterogeneous
records from the health system, mortality because the Colombian armed conflict and administrative data from the education system. For the proposed procedure the data is obtained with the integrated administrative record and prior information can be
introduced for the different age groups and genders using both the expertise of the practitioner and
the population projections computed with the cohort-component method. Let $Q_{l,t}$ be a demographic interest quantity obtained for instance through the number of births, deaths, migrants and habitants in the different demographic breakdowns of age, sex and time period.\\

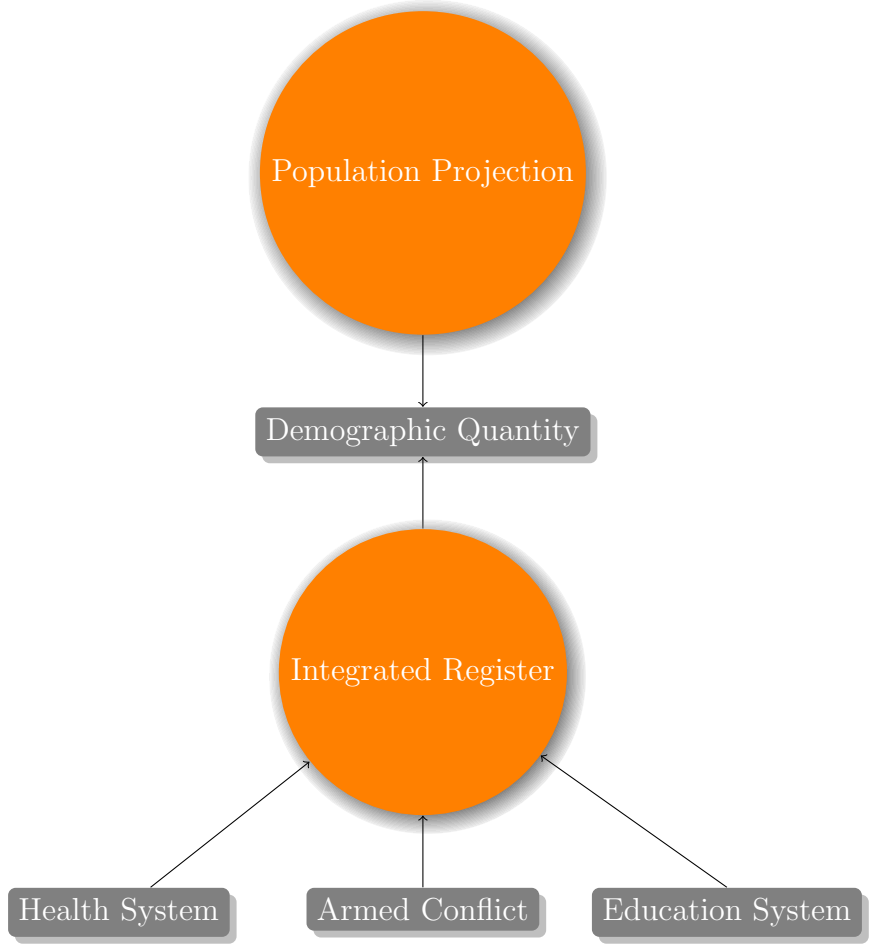
\begin{figure}[ht]
\begin{center}
\begin{tikzpicture}[
    fact/.style={rectangle, draw=none, rounded corners=1mm, fill=gray, drop shadow,
        text centered, anchor=north, text=white},
    state/.style={circle, draw=none, fill=orange, circular drop shadow,
        text centered, anchor=north, text=white},
    leaf/.style={circle, draw=none, fill=red, circular drop shadow,
        text centered, anchor=north, text=white},
    level distance=0.95cm, growth parent anchor=south
]
\node (State00) [state] {Population Projection} [->]
    child{
        node (Fact01) [fact] {Demographic Quantity}[->]
        child{ [sibling distance=4cm]
            node (State01) [state] {Integrated Register}[<-]
            child{
             node (Fact02) [fact] {Health System}
            }
                        child{
             node (Fact02) [fact] {Armed Conflict}
            }
            child{ [sibling distance=2cm]
                node (Fact10) [fact] {Education System}
            }
        }
    }
;

\end{tikzpicture}
\end{center}
\caption{{\textcolor{blue}{
Representation of the proposed procedure.}}}
\label{fig:modelo}

\vspace{0.5cm}

\end{figure}

In practice $Q_{l,t}$ is projected by $q^{c}_{l,t}$ using for example the cohort-component method which may be biased
for long intercensal periods.  We consider $M_{Q}$ a model to represent the variable $Q$ with variables obtained through the compensation equation which includes births, deaths, immigration and emigration. In practice we can have $K$ heterogeneous administrative records with $X_{1},...,X_{K}$ variables which can used to model the quantity $Q$ using for instance regression models, time series or through of direct estimations. Figure \ref{fig:modelo} illustrates the proposed model where the considered administrative records are obtained from 1) the health system using: a) Beneficiaries of Social Programs (SISBEN, by its Spanish acronym), b) Unique Affiliates Database (BDUA, by its Spanish acronym), c)
the NUEVA EPS (Healthcare Promoting Entity), d) the Sanitas EPS and e) the Births and Deaths Registry; 2) the mortality due to the Colombian armed conflict using
the Single Victims Registry (RUV, by its Spanish acronym) and 3) the administrative data from the education system using the Student Enrollment System for Basic and Middle Education (SIMAT, by its Spanish acronym).\\

{\textcolor{blue}{The proposed procedure given in Figure \ref{fig:modelo} 
is different to other proposals  (e.g., \citep{bryant2013bayesian, bryant2015bayesian}) 
where there is a model for each administrative record and therefore the
uncertainty becomes from both the sum of uncertainties from the models of the administrative records and the uncertainty from the model of the
demographic quantity (in a hierarchical Bayesian structure).  
We also consider other important administrative records because the characteristics of some developing countries where 
there is mortality because the armed conflict. Additionally, our focus is improving the quality of population projections
by building base population estimates for countries where there are long intercensal periods as is the case of some
Latin American countries.  Therefore we use a deterministic methodology proposed in Chapter 5 of \cite{wallgren2007register} to build a base population integrated administrative record using data from different heterogenous sources. This integrated record is produced by making an inventory and defining, processing and integrating the information into a single administrative record. Therefore the uncertainty becomes only from the Bayesian model (see the narrow credible intervals in the next section Figure \ref{fig:proyecta}, top right) and it does not depend of the uncertainty obtained by considering a model for each administrative record.}}\\


The quality of the database register is contingent upon the quality of the administrative records. Although in this work we build the integrated record using deterministic
methods we could consider probabilistic methods given by \cite{christen2005probabilistic,herzog2007data}. Table \ref{reg} shows a summary from each of the administrative records and the migrations of these records to another system as well the duplicates and deaths. We found that the number of duplicates
were 805  highlighting the significant contribution of the records in the construction of a base population. Additionally the percentage of migration and deaths is
very small (7.58\%) according to Table \ref{reg}.

\begin{table}[ht]
\caption{Results of linking the administrative records using deterministic methods.}
\centering
\begin{tabular}{ccccccccccccc}
  \hline   \hline
	States	& Migration	& Duplicate	& Death	& Contribution\\ \hline   \hline
SISBEN  & 1171 & 491 & 563 & 74449\\
Deaths & 0 & 0 & 5821 & 0\\
SIMAT & 40 & 176 & 4 & 7879\\
Nueva EPS & 183 & 0 & 46 & 5868\\
Sanitas EPS & 46 & 0 & 0 & 901\\
RUT & 684 & 45 & 25 & 13996 \\
Births & 80 & 28 & 1 & 3264\\
RUV & 109 & 0 & 23 & 4485\\
BDUA  & 1168 & 65 & 249 & 30794\\   \hline   \hline
Total & 4081 & 805 & 6732 & 142636 \\
  \hline   \hline
\end{tabular}
\label{reg}
\end{table}

\begin{figure}[ht]
\begin{center}
\begin{tikzpicture}[
    fact/.style={rectangle, draw=none, rounded corners=1mm, fill=gray, drop shadow,
        text centered, anchor=north, text=white},
    state/.style={circle, draw=none, fill=orange, circular drop shadow,
        text centered, anchor=north, text=white},
    leaf/.style={circle, draw=none, fill=red, circular drop shadow,
        text centered, anchor=north, text=white},
    level distance=0.95cm, growth parent anchor=south
]
\node (State00) [state] {$\boldsymbol{\alpha}$} [->]
    child{
        node (Fact01) [fact] {$\boldsymbol{\theta}$}[->]
        child{ [sibling distance=4cm]
            node (State01) [state] {$\boldsymbol{q}=f(X_{1},...,X_{K})$}[<-]
            child{
             node (Fact02) [fact] {$X_{1}$}
            }
                        child{
             node (Fact02) [fact] {$X_{2}$}
            }
            child{ [sibling distance=2cm]
                node (Fact10) [fact] {$X_{K}$}
            }
        }
    }
;

\end{tikzpicture}
\end{center}
\caption{{\textcolor{blue}{
Representation of the proposed Bayesian approach.}}}
\label{fig:modelo22}
\end{figure}
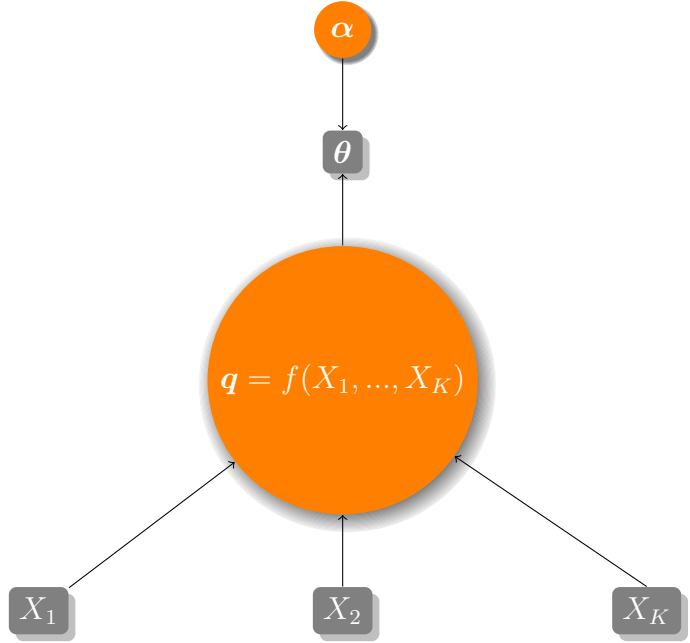

We consider now the proposed model. Let be $\boldsymbol{q}=f(X_{1},...,X_{K})$ a vector with the information of the interest quantity given by
the integrated record and where  $f(X_{1},...,X_{K})$ is a function of the $X_{1},...,X_{K}$ administrative records.
The proposed procedure based in a Bayesian approach is useful to compute estimates of base populations using an integrated record for a specific period of time. The Bayesian model is given by

\begin{align}\label{mult}
  &\boldsymbol{q}\sim \text{Multinomial}(\boldsymbol{\theta},n) \\
  &\boldsymbol{\theta} \sim \text{Dirichlet}(\boldsymbol{\alpha}) \notag
\end{align}

where $\boldsymbol{q}={q_{1,t},...,q_{L,t}}$, $\boldsymbol{\alpha}=\{\alpha_{1,t},...,\alpha_{L,t}\}$ and $\boldsymbol{\theta}={\theta_{1,t},...,\theta_{L,t}}$ with $L$
the number of the age-groups and where $n$ the total population size for the considered sex-group. The considered model is illustrated in Figure \ref{fig:modelo22}.
The posterior distribution  of using (\ref{mult}) is Dirichlet  (see for instance \cite{gelman2013bayesian}) as follow

\begin{align}\label{dir2}
p(\boldsymbol{\theta}|\boldsymbol{\alpha^{*}})=\dfrac{1}{\mathbf{B}(\boldsymbol{\alpha^{*}})}
\prod_{l=1}^{L}\theta_{l,t}^{\alpha^{*}_{l,t}-1},
\end{align}

where $\alpha^{*}_{l,t}=q_{1,t}+\alpha_{1,t}$, $\sum_{l=1}^{L}\theta_{l,t}=1$ and $\mathbf{B}(\boldsymbol{\alpha^{*}})$ the normalization constant computed
by using gamma functions as follow

\begin{align}
\mathbf{B}(\boldsymbol{\alpha^{*}})= \dfrac{\prod_{l=1}^{L}\Gamma(\alpha^{*}_{l})}{\Gamma(\Sigma_{l=1}^{L}\alpha^{*}_{l})}.
\end{align}

 The value of $\boldsymbol{\alpha}=\{\alpha_{1,t},...,\alpha_{L,t}\}$
 represents the number of observations in each age group $l=1,...,L$ for time $t$ which have been computed previously using the cohort-component method. $\boldsymbol{\alpha}$  is also useful in practical terms if we wish to include additional information on an age-group $l$
 through the demographer criteria. An important practical advantage is due to the fact that the posterior distribution given in (\ref{dir2}) is computed in closer form and credibility intervals for age-groups estimates in the base population can be constructed using the quantiles of the Dirichlet distribution.

\section{Population base estimates for the municipality of Jamund\'i}
\label{ssec:tres}

\begin{figure}[ht]
\begin{center}
\begin{tabular}{ccc}
\includegraphics[scale=0.75]{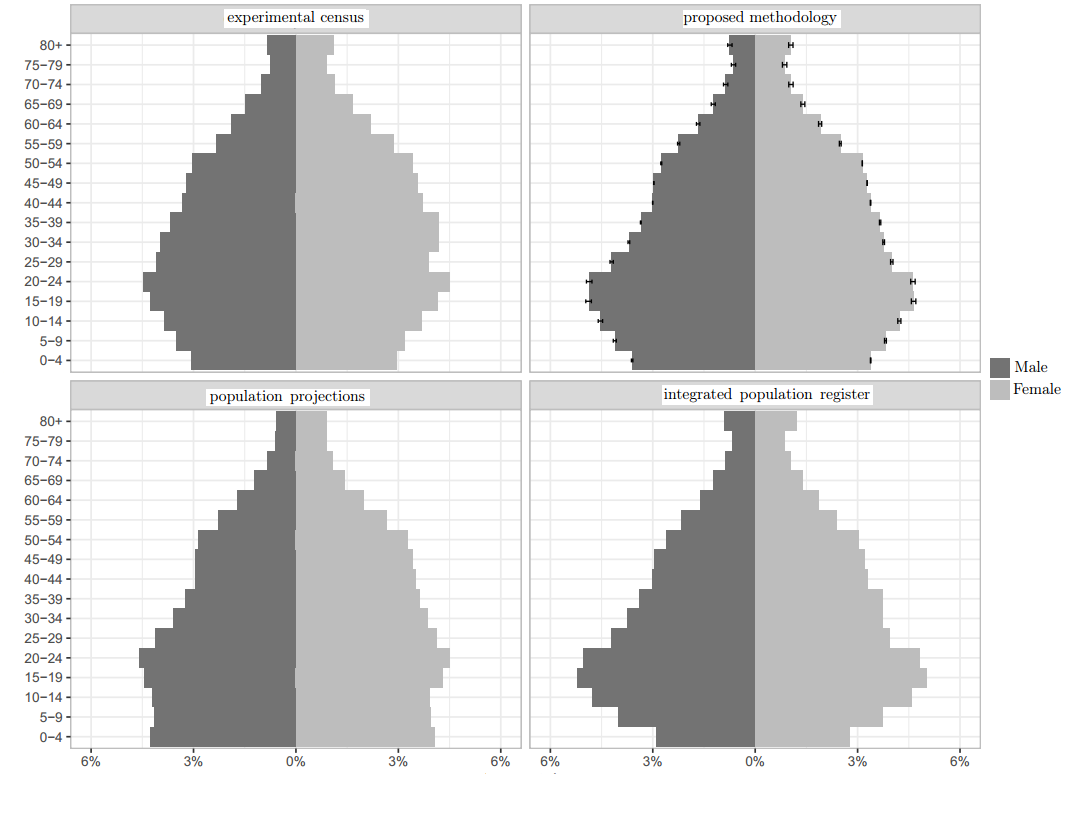}
\end{tabular}
\end{center}
\caption{Estimates of base population in the municipality of Jamundí in 2016.  Using the experimental census (top-left),
the proposed methodology (top-right), the official population projections of DANE (bottom-left), and the integrated population register (bottom-right).}
\label{fig:proyecta}
\end{figure}

In order to illustrate the advantages of the proposed methodology we compute the base population estimates for the municipality of Jamundí in Colombia for 2016.
Jamund\'i is located in the department of Valle del Cauca in Colombia which has been affected by the Colombian armed conflict and where
live an indigenous group called the p\'aez, also known as the Nasa.  We assess if the population estimates are able to capture the size population structure by age and sex using an experimental census developed in the municipality of Jamund\'i in 2016. Figure \ref{fig:proyecta} illustrates the population structures according to age and sex obtained through: i.) The experimental census (used as a benchmark to compare the different estimates); ii.) The proposed methodology; iii.) The official population projections of DANE; and iv.) The integrated population register discussed in Section \ref{ssec:dos}.\\

According to the results in Figure \ref{fig:proyecta} the base population estimates through the Bayesian model shows a population structure similar to the one obtained through of the experimental census data. The proposed approach produces an  intermediate pattern between the one projected by the cohort relation method which has a greater relative weight of the early-age population and the estimate of the integrated population record where the early-age population is unrepresented. The population dynamics did not behave as expected in the projections may be due to the fact that the decline in fertility appeared earlier and was more accelerated than the expected. Moreover there may be a lack of coverage in the posterior distribution (see Figure \ref{fig:proyecta}, top-left) where the infant population in the administrative records may be is unrepresented. This  may be improved by incorporating in the integrated population register the information of vaccines  for the five-age population computed by the ministry of health in Colombia. \\

We now use the Absolute Percentage Differences (APD) in order to compare the estimates of using our proposal, the cohort relation method or the integrated record with respect to the  micro-data in the experimental census for each combination of age and sex.
Tables \ref{table1} and \ref{table2} summarized the APD averages for age and sex respectively. The most interesting remark of this analysis is the improvement obtained in the precision of population estimates using the proposal in almost all age groups and for both sexes. Particularly by using the proposed Bayesian model all population estimates by age (Table \ref{table1}) show an APD below 20 percent and in most cases below 15 percent.\\

\begin{table}[ht]
\caption{Averages of Absolute Percentage Differences in terms of the Experimental Census according to age groups using the
proposed methodology, the official population projections and the integrated population register.}
\centering
\begin{tabular}{rlrrr}
    \hline   \hline
& Age	& Proposed Methodology & Official Projections & Integrated Register\\
    \hline   \hline
1 & 0-4 & 13.66 & 92.83 & 34.76 \\
  2 & 5-9 & 15.49 & 68.24 & 64.70 \\
  3 & 10-14 & 13.29 & 49.56 & 76.97 \\
  4 & 15-19 & 10.31 & 44.76 & 73.20 \\
  5 & 20-24 & 3.10 & 40.85 & 56.28 \\
  6 & 25-29 & 0.21 & 43.34 & 45.51 \\
  7 & 30-34 & 10.87 & 26.45 & 30.43 \\
  8 & 35-39 & 12.78 & 21.70 & 29.68 \\
  9 & 40-44 & 11.67 & 27.12 & 27.48 \\
  10 & 45-49 & 9.85 & 30.60 & 29.25 \\
  11 & 50-54 & 11.12 & 32.00 & 24.22 \\
  12 & 55-59 & 10.49 & 32.36 & 25.59 \\
  13 & 60-64 & 14.69 & 24.72 & 21.18 \\
  14 & 65-69 & 18.36 & 16.48 & 18.88 \\
  15 & 70-74 & 11.93 & 25.51 & 28.37 \\
  16 & 75-79 & 10.60 & 23.87 & 33.89 \\
  17 & 80+ & 9.76 & 7.56 & 55.86 \\
     \hline   \hline
\end{tabular}
\label{table1}
\end{table}

In contrast, all the estimates between 0 and 79 years obtained in the official projections show APD above 15 percent and particulary the estimates obtained in early ages were quite imprecise. In the case of the integrated record, no APD per age presented values lower than 15 percent. Furthermore for the comparison of estimates per sex (Table \ref{table2}) there is remarkable advantage of using the proposed methodology because the average of the APDs are more or less equivalent to a third of the ones obtained through the other two contrasted methodologies.

\begin{table}[ht]
\caption{Averages of Absolute Percentage Differences in terms of the Experimental Census according to sexes using the
proposed methodology, the official population projections and the integrated population register.}
\centering
\begin{tabular}{rlrrr}
  \hline   \hline
 & Sex	& Proposed Methodology & Official Projections & Integrated Register\\    \hline   \hline
1 & Male & 11.55 & 33.90 & 40.48 \\
  2 & Female & 10.62 & 37.69 & 39.10 \\
  3 & Total & 11.03 & 35.69 & 39.76 \\
  \hline   \hline
\end{tabular}
\label{table2}
\end{table}

\section{Concluding remarks and  future work}
\label{ssec:cuatro}


In this paper a practical Bayesian procedure is proposed to estimate base populations for long intercensal periods  in small areas through previously integrated information from heterogeneous administrative records. The proposal is robust in the estimation of population pyramids particularly in small areas where the coverage of the administrative records is high. The prior information
in the proposed model can be introduced by both using the traditional components method and the expertise of
the practitioner. The proposed model is robust to estimate population base as is illustrated in the municipality of Jamund\'i in Colombia. We found the proposed alternative very promising to develop and update population structures on a local level in small areas and we now have some suggestions on future work: 1) Because in the integrated population register
the early-age population is unrepresented we could explore the inclusion of a new administrative
record computed using the vaccine information of the early-age population, 2) We  observed some preliminary advantages
of applying Bayesian dynamic models in the Unique Affiliates Database for temporal
population base estimates. Therefore, a possible improvement in the proposed model is to incorporate some temporal behaviour, and 3) In this work we consider a Dirichlet prior for the base population however in practice this is may be a strong assumption because for
the age-groups different administrative sources are considered. Therefore a more flexible prior may be considered using a
prior based in a Dirichlet process mixture \citep{hjort2010bayesian}.

\section*{Acknowledgements}

This project was supported in part by the Bloomberg Philanthropies Data for Health Initiative,
National Administrative Department of Statistics and Vital Strategies.  We thank Adrian E. Raftery (University of Washington) for his valuable comments  on this paper.
We also thank the panel participants in the workshop on Small Area Demographic Estimation in Latin America (Rio de Janeiro, Brazil 2018) for
the discussion and suggestions on this approach and the technical committee of DANE.

\bibliographystyle{plainnat}

\bibliography{bibliosmallb}            

\end{document}